\documentclass[twocolumn,trackchanges]{aastex631}
\usepackage{amsmath}
\usepackage{epsf}
\usepackage{color}
\usepackage{color}
\usepackage{enumitem}
\usepackage{mathtools}
\usepackage{svg}
\usepackage{booktabs}
\usepackage[mathscr]{euscript}
 \let\mathscr\relax
\usepackage[scr]{rsfso}

\shorttitle{\texttt{NotPlaNET}}
\shortauthors{tardugno poleo, eisner, and hogg}

\turnoffedit 

\begin{document}
\title{\texttt{NotPlaNET}: Removing False Positives from \textsl{Planet Hunters TESS} with Machine Learning}

\author[0009-0002-3883-7490]{Valentina Tardugno Poleo}
\affiliation{Center for Cosmology and Particle Physics, Department of Physics, New York University, 726 Broadway, New York, NY 10003, USA}
\email{vt2189@nyu.edu}

\author[0000-0002-9138-9028]{Nora Eisner}
\affiliation{Center for Computational Astrophysics, Flatiron Institute, 162 Fifth Avenue, New York, NY 10010, USA}
\author[0000-0003-2866-9403]{David W. Hogg}
\affiliation{Center for Cosmology and Particle Physics, Department of Physics, New York University, 726 Broadway, New York, NY 10003, USA}
\affiliation{Center for Computational Astrophysics, Flatiron Institute, 162 Fifth Avenue, New York, NY 10010, USA}
\affiliation{Max-Planck-Institut für Astronomie, Königstuhl 17, D-69117 Heidelberg, Germany}

\begin{abstract}\noindent
Differentiating between real transit events and false positive signals in photometric time series data is a bottleneck in the identification of transiting exoplanets, particularly long-period planets. This differentiation typically requires visual inspection of a large number of transit-like signals to rule out instrumental and astrophysical false positives that mimic planetary transit signals. We build a one-dimensional convolutional neural network (CNN) to separate eclipsing binaries and other false positives from potential planet candidates, reducing the number of light curves that require human vetting. Our CNN is trained using the \textsl{TESS} light curves that were identified by \textsl{Planet Hunters} citizen scientists as likely containing a transit. We also include the background flux and centroid information. The light curves are visually inspected and labeled by project scientists and are minimally pre-processed, with only normalization and data augmentation taking place before training. The median percentage of contaminants flagged across the test sectors is 18\% with a maximum of 37\% and a minimum of 10\%. Our model keeps 100 \% of the planets for 16 of the 18 test sectors, while incorrectly flagging one planet candidate (0.3\%) for one sector and two (0.6\%) for the remaining sector. Our method shows potential to reduce the number of light curves requiring manual vetting by up to a third with minimal misclassification of planet candidates. 
\end{abstract}



\section{Introduction}\label{sec:intro}

There have been significant advancements in space-based, precise time-series measurements, particularly with missions such as \textsl{Kepler} \citep{kepler} and the Transiting Exoplanet Survey Satellite \citep[\textsl{TESS}; ][]{tess}. These advances have led to large numbers of exoplanet discoveries. To effectively parse through the growing volume of data, the field requires innovative analysis techniques. 

Despite the growing selection of transit-finding and classifier models, the process of identifying transiting exoplanets still relies heavily on human vetting. One of the most successful methods used to identify periodic, transit-like signals in photometric data is the Box Least Squares \citep[BLS; ][]{bls2002} algorithm. However, this method typically produces more candidates than true planets, with a large fraction of the identified signals being false positives that mimic true transits. These false positive scenarios include eclipsing binaries and systematic variations caused by the instrumentation, cosmic rays, or temperature fluctuations. Typically, teams of experts examine possible transit signals and assess their disposition (e.g. \citealt{Guerrero_2021, 2018Crossfield, toi813}). This process is very time-consuming, particularly if each candidate is inspected by multiple experts. Other vetting methods consist of fitting physical and probabilistic models to light curves. \citet{dfm} present a fully automated model that operates without human intervention. Yet the described method struggles to identify long-period planets orbiting variable stars.

Convolutional Neural Networks \citep[CNNs;][]{cnn}, a widely used machine learning tool for image pattern recognition, have also been applied to help the classification process. Progress has been made in employing CNNs and other machine learning techniques to identify light curves that contain transit-like signals as well as to classify these signals as planet candidates or false positives (\citealt{Shallue_2018}, \citealt{Yu}, \citealt{Tey_2023}, \citealt{malik2022}, \citealt{Dattilo_2019}). Nevertheless, human vetting still significantly outperforms current models when identifying long-period, single-transit, and aperiodic events (e.g. \citealt{wang}, \citealt{circumbinary}, \citealt{ph}).

Typically, machine-learning classification methods rely on the use of phase-folded time-series data and thus require prior knowledge of the periodicity of the signal \citep[e.g. ][]{2023Salinas}. The requirement of two transit events for phase-folding also restricts their use to multi-transit planets and thus shorter period planets. Furthermore, most of these methods do not consider background or centroid information, which may help identify background events and eclipsing binaries. \textcolor{black}{Other CNNs make use of heavily pre-processed data.} A more flexible model that does not require phase-folding or extensive pre-processing, and that considers background and centroid information could thus complement existing models.



We built and trained a one-dimensional CNN to remove as many false positive signals as possible without discarding planet candidates. \textcolor{black}{In contrast to previous work, \citep[e.g., ][]{2023Salinas, 2021Olmschenk, 2018Zucker}, our model focuses on contaminant detection and does not rely upon the use of known orbital periods.} The open-source CNN, \texttt{NotPlaNET}\footnote{\url{https://github.com/vtardugno/TESS-CNN}}, was trained using labels provided by the \textsl{Planet Hunters TESS} citizen science project. The labels consist of times within light curves where a transit-like signal has been flagged, and the associated label of `likely planet candidate' (PC), `eclipsing binary' (EB), or `other' (OTH). Since the CNN was purposely designed to work with data that has not been phase folded, \texttt{NotPlaNET} allows us to vet single-event signals and help identify single-transit, longer-period planets. The vetting and identification of single-transit signals is particularly important in the era of \textsl{TESS}, where the majority of planets with periods longer than 27 days will not exhibit consecutive transits in the \textsl{TESS} data. 


The paper is structured as follows: Section \ref{sec:data} describes the data and pre-processing that was carried out. Section \ref{sec:cnn} provides an overview of the neural network's architecture. The contaminant selection procedure and the results are outlined in Section \ref{sec:class}. Section \ref{sec:disc} provides a discussion of the results, and lastly, Section \ref{sec:conc} concludes with a summary of our results and directions for future research. 

\section{Data, labels, and preprocessing}\label{sec:data}

The data were composed of the two-minute cadence \textsl{TESS} \textcolor{black}{Simple Aperture Photometry (SAP) and Pre-search Data Conditioning SAP (PDCSAP)  light curves}, which are produced by the \textsl{TESS} pipeline at the Science Processing Operations Center (SPOC) \citep{spoc} and made publicly available by the \textcolor{black}{Mikulski Archive for Space Telescopes} (MAST). Each sector observed by \textsl{TESS} measures $24^{\circ} \times 96^{\circ}$ oriented along the line of ecliptic longitude and consists of $\sim$ 27 days of photometric observations. The data also included eclipsing binaries from the \citet{ebs} \textsl{TESS} Eclipsing Binary Stars catalog and \textcolor{black}{Tess Objects of Interest \citep[TOI;][]{toi}}. We used data from sectors 2-6, 10-11, 17-19, and 32-60. Sectors ranging from 2 to 35 were used to train the network. The training sectors were randomly selected from the earliest 35 sectors to assess whether a model trained on early sectors could have predictive power for unseen newer sectors. We left gaps in the selected sectors to ensure our model could generalize to unseen sectors. Sectors 36-42 were used to optimize our model's hyperparameters and fine-tune our final classification scheme. Sectors 43-60 were left unused until the very end of the evaluating phase so that we could assess how well our model performed on data it had not been trained or validated on.  

\subsection{Labels}
\label{sec:pht}

We make use of the light curve labels from the \textsl{Planet Hunters TESS} citizen science project \cite[for details see][]{eisner2020method}. In brief, the online project presents volunteers with individual \textsl{TESS} light curves and asks them to identify the times of possible transit-like signals using their mouse button. Each light curve is seen by 15 volunteers, who independently identify transit signals. For each light curve, the markings from the individual volunteers are combined using a density-based clustering algorithm, that allows us to identify light curves where multiple volunteers identified the same signal. Taking into consideration the number of volunteers who identified a given signal and their user weights \cite[see][]{eisner2020method} we calculate a transit score for each \textsl{TESS} light curve for a given sector (a higher score indicates a higher confidence in the signal being real). 

For each \textsl{TESS} sector, the light curves with the 500 highest ranked scores are visually inspected by three members of the \textsl{Planet Hunters TESS} science team to differentiate between possible planet candidates (PC), eclipsing binaries (EB) and \textcolor{black}{other signals (OS)}. This manual vetting of the light curves has been completed for sectors 1-65. 

It is important to note that while we relied on the scientists' labels as ground truth during the training and validation of our model, there remains a degree of uncertainty in these labels. In particular, the scientists are encouraged to vet light curves `optimistically', meaning that if the nature of the signal is ambiguous, it should be classified as a PC. Identifying ambiguous signals as potential planets allows for these light curves to be inspected in the next round of vetting, minimizing the loss of potential candidates, but this can lead to the mislabeling of several targets.

\textcolor{black}{Notably, while some of the light curves and their labels were taken from the \textsl{TESS} EB catalog and TOIs, the exact locations of the transits were derived from the volunteer markings in \textsl{Planet Hunters TESS}. This ensured that the methodology for determining signal locations was consistent across planets, EBs, and other signals, with all signal locations being determined via \textsl{Planet Hunters TESS}.}

\subsection{Preprocessing}

All of the \textsl{TESS} light curves with a PC, EB, or \textcolor{black}{OS} label were \textcolor{black}{first} normalized and \textcolor{black}{then} split into smaller chunks for data augmentation purposes as well as to ensure translational invariance. The light curves and accompanying data, except for the background flux, were normalized by dividing by the median value of each array. Conversely, the background flux was normalized by applying a median filter to correct for the time-variable scattered light from the Earth and the Moon. Then, all the data were parsed using a stride of 800 data points and chunked into 2-day fragments. \textcolor{black}{The choice of a 2-day fragment size stems from our main objective: to avoid discarding planets, especially single-transit and long-period ones. Given that such planets may have orbital periods spanning several days, we opted for a fragment size capable of capturing these extended transit events.} 

\textcolor{black}{Gaps in the light curves were handled by appending the data as if the gaps did not exist. This approach was chosen because our focus is on single events within narrow windows, rather than periodic signals, making the influence of these gaps minimal on our training process. Interpolating over large data gaps was avoided to prevent introducing bias and distorting the signal characteristics. It remains an open question how to best handle data gaps, and further research is required to explore techniques to best address the problem of missing data.}

\textcolor{black}{To assign the labels to each chunk, we examined whether each fragment contained a citizen scientist marking. If so, the fragment was assigned the label given during manual vetting. If not, the fragment was assigned an OS label.} To create training and validation sets, the data were shuffled and split using sklearn's \citep{scikit} train\_test\_split function with its default settings.

\section{Neural Network Architecture}\label{sec:cnn}

\begin{figure}[h!]
\centering
\includegraphics[width=0.4\textwidth]{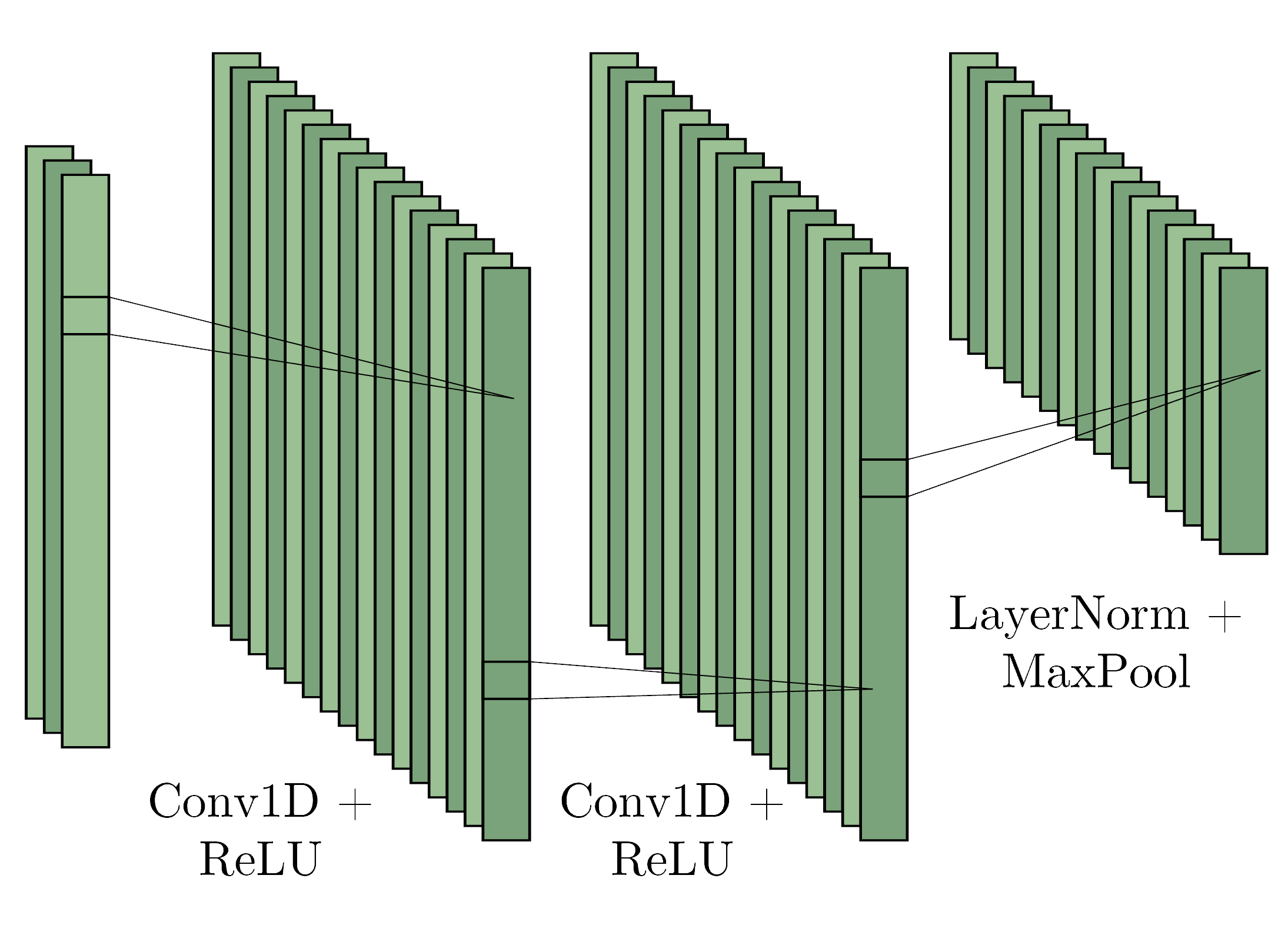} 
\caption{Convolutional block diagram. The block consists of two 1D convolutional layers with the ReLU activation function, layer normalization, and max pooling.}
\label{fig:cnn} 
\end{figure}

The CNN was built with PyTorch \citep{pytorch} and consists of six 1D convolutional blocks. Each convolutional block has two 1D convolutions with a kernel size of 5, followed by the ReLU activation function, layer normalization, and max pooling. Lastly, we added one linear layer with the Softmax activation function. Figure~\ref{fig:cnn} illustrates a single convolutional block, and Table~\ref{tab:cnn_layers} describes the entire architecture. The architecture was chosen using hyperparameter optimization with the Weights and Biases development platform \cite{wandb}. The network's inputs were the light curve chunks along with their background fluxes and centroids, and the outputs consisted of three numbers which were scores assigned to each category.

\begin{table*}[t]

\begin{tabular}{lccccc} 
\toprule
\textbf{Layer} & \textbf{Output Size (LxC)}  & \textbf{Activation} & \textbf{Normalization} & \textbf{Pooling} & Dropout \\ \midrule
Conv1\_1/Conv1\_2 & 1440x32 & ReLU & Layer Norm & Max Pooling & 10\%\\
Conv2\_1/Conv2\_2 & 720x64 & ReLU & Layer Norm & Max Pooling & 10\%\\
Conv3\_1/Conv3\_2 & 360x128  & ReLU & Layer Norm & Max Pooling & 10\%\\
Conv4\_1C/onv4\_2 & 180x256  & ReLU & Layer Norm & Max Pooling & 10\%\\
Conv5\_1/Conv5\_2 & 90x64  & ReLU & Layer Norm & Max Pooling & 10\%\\
Conv6\_1 & 45x32  & ReLU & - & - & -\\
Conv6\_2 & 45x16  & ReLU & Layer Norm & Max Pooling & -\\ 
Flatten & 16x1 &-&-&-&-\\
Linear & 16x3 & Softmax &-&-&-\\
\bottomrule
\end{tabular}
\caption{\texttt{NotPlaNET} Convolutional Neural Network parameters. All values were selected through hyperparameter optimization by measuring which configurations yielded the lowest validation loss.}
\label{tab:cnn_layers}
\end{table*}

The Adam optimizer \citep{adam_opt} with a learning rate of 0.0005 was used to minimize the cross-entropy loss. Due to the imbalanced nature of the data, with EBs being the most represented, we assigned loss weights to each category. A learning rate scheduler was set to decrease the learning rate by a factor of 0.8 every 50 epochs and the network was trained for 200 epochs.  During every training loop, 10\% of the nodes at the end of each convolutional block were randomly zeroed, leading to a different configuration after every epoch. Dropout regularization helps the network learn patterns in the data rather than memorizing the training set, avoiding overfitting. In every epoch the training and validation loss were calculated and the weights associated with the lowest validation loss were stored as our final model parameters.

\section{Classification Methods and Results}\label{sec:class}

\begin{figure}[h!]
\centering 
\includegraphics[width=0.45\textwidth]{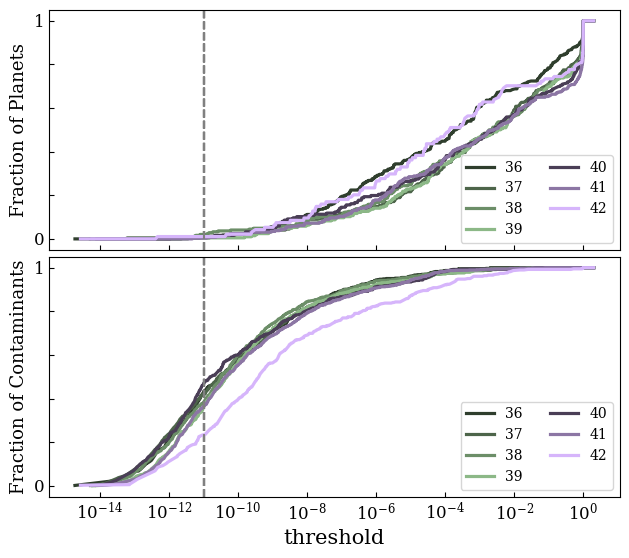}  
\caption{Fraction of planet candidates (top) and true contaminants (bottom) flagged as contaminants as a function of the PC score threshold for the seven validation sectors. For each validation sector, we chose a threshold that maximized the fraction of discarded contaminants while discarding zero planets. We then chose our final threshold (dashed line) by performing a weighted average of each validation sector's threshold. The weights were assigned as the inverse square of the number of contaminants found in a given sector, allowing for a more conservative cutoff.  }
\label{fig:roc} 
\end{figure}

We devised a selection process that used the CNN scores to determine whether a given light curve should be flagged as a contaminant. We aimed to find a threshold that found as many contaminants as possible while removing no planet candidates in the validation sectors. For each validation sector, we probed CNN prediction thresholds in all three categories and found that the best results were achieved when placing a threshold on the PC score. Figure \ref{fig:roc} illustrates the effect of different PC thresholds on the fraction of planet candidates and true contaminants discarded.

The threshold for each validation sector was chosen so that no planet candidates were discarded and the number of true contaminants found was maximized. We then found the average of the sectors' thresholds, weighted by the inverse square of the number of contaminants found per sector. The weight equation was chosen empirically, using the validation sectors to identify what final threshold resulted in the highest combined fraction of true contaminants found while imposing the condition that no planet candidate is incorrectly classified. We suspect the reason behind the success of this weighting scheme was that it resulted in lower weights for thresholds that discarded more contaminants, yielding a more conservative cut-off.

\begin{figure*}[t]
\centering
\includegraphics[width=1\textwidth]{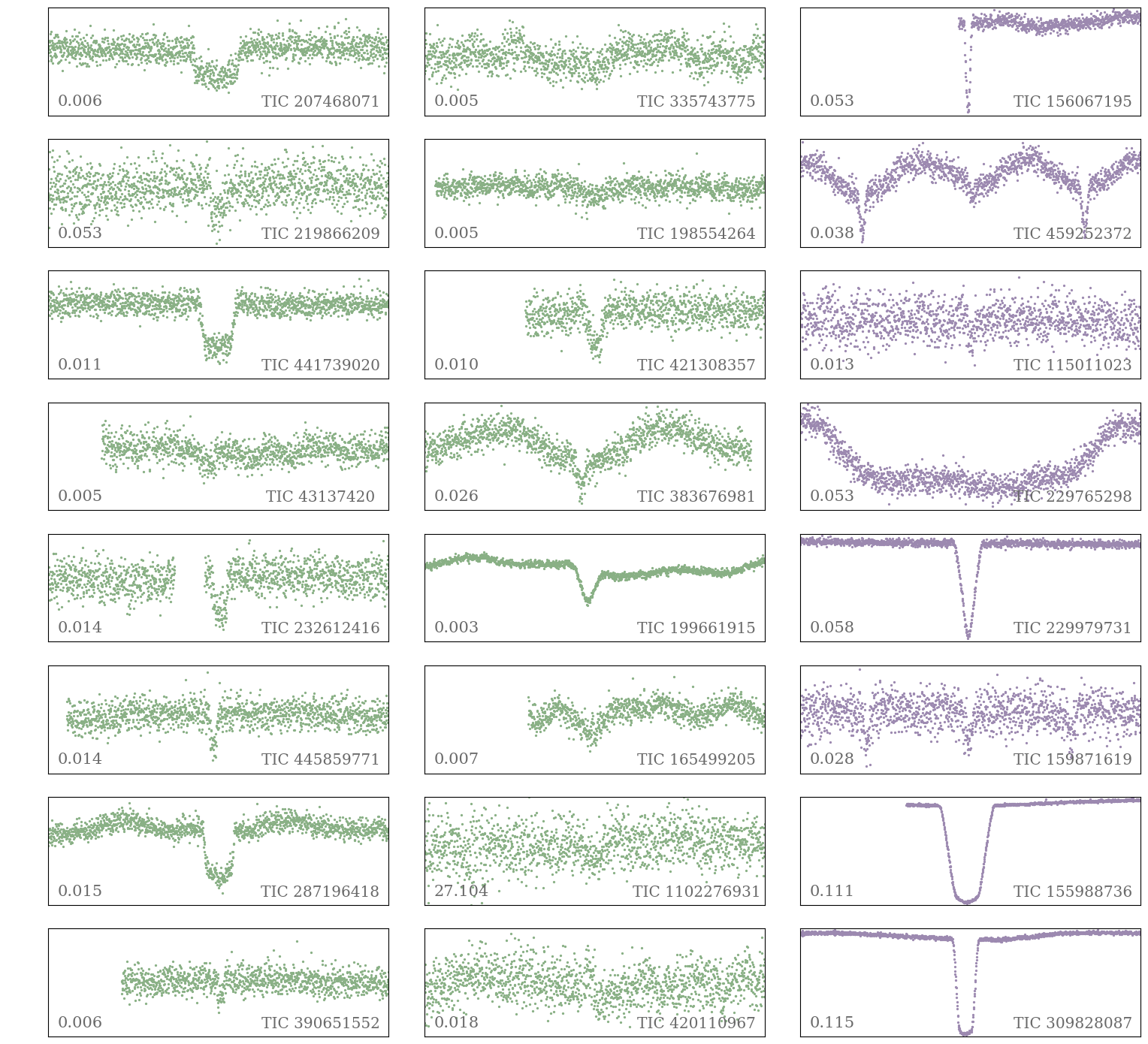}
\caption{\textcolor{black}{Selected light curve chunks from sector 50. The rightmost column (purple) contains the light curves flagged as contaminants by our model. The leftmost and middle columns (green) contain light curves classified as `keep for further vetting.' The left column shows true planet candidates, while the middle one shows true contaminants that were incorrectly classified as `keep for further vetting.' The y-axis range is displayed in the lower-left corner of each panel, while the lower-right corner displays the TIC ID of the light curve.}}
\label{fig:tshirt} 
\end{figure*}

Before making a classification, we looked at all the scores assigned to each chunk of a given TIC ID and sector. Light curves with a PC score lower than the determined threshold in \textsl{all chunks} were labeled as contaminants. As long as at least one chunk in the light curve had a PC score greater than the threshold, the light curve was assigned to the `keep for further inspection' category. Figure \ref{fig:tshirt} displays selected light curve chunks for the `contaminants' and `keep for further vetting' categories. Although we selected a very conservative threshold, the model can easily be modified to more aggressively reduce contamination at the expense of a few missed planet candidates. 

We applied \texttt{NotPlaNET} along with our chosen classification method and threshold on the test set, containing data from sectors that had not been previously used for training, threshold selection, or any other decision-making steps. The classification results are summarized in Table \ref{tab:results}

\begin{table}[h]
    \centering
    \begin{tabular}{ccc|ccc}
       Sec & PC & EB+\textcolor{black}{OS} (\%) & Sec &PC & EB+\textcolor{black}{OS} (\%)  \\ \hline
        43 & 0 & 68 (11\%) & 52 & 0 &128 (18\%) \\
        44 & 0 &115  (17\%) & 53 & 2 &162 (20\%) \\
        45 & 0 & 73 (17\%)   & 54 & 0 & 10 (20\%) \\
        46 & 0 & 15 (20\%)   & 55 & 0 & 46 (14\%) \\
        47 & 0 &125 (16\%)   & 56 & 0 &175 (22\%) \\
        48 & 0 &124 (16\%)   & 57 & 0 &113 (22\%) \\
        49 & 0 & 92  (12\%)  &  58 & 1 &288 (36\%) \\
        50 & 0 & 82  (10\%)  &  59 & 0 & 89 (19\%)\\
        51 & 0 & 43 (17\%)  &   60 & 0 & 55 (15\%)\\
        
    \end{tabular}
    \caption{Light curves flagged as contaminants by our model. The planets wrongly classified as contaminants are listed on the PC column. The EB+\textcolor{black}{OS} column lists all the true contaminants discarded in each sector along with the percentage of total contaminants discarded.}
    \label{tab:results}
\end{table}

For our test set, the percentage of EBs in the found contaminants had a median value of 82\% across all test sectors with a maximum of 100\% and a minimum of 68\%. However, the median, maximum, and minimum percentages of EBs discarded out of the total number of EBs per sector were 23\%, 44\%, and 16\%, respectively. While our model cannot yet find a high fraction of the EBs in the data, a high fraction of the flagged contaminants we found were EBs. In other words, the contaminant category has high precision and low recall rates when looking for eclipsing binaries. We suspect our model could be modified and applied to projects focusing on eclipsing binary stars. \textcolor{black}{Moreover, we found no significant difference in the morphology distributions of eclipsing binaries that were labeled as contaminants and the ones that were kept for further vetting. }

\

\bigbreak

\

\section{Discussion}\label{sec:disc}

\texttt{NotPlaNET} demonstrated the potential to identify \textcolor{black}{contaminants} while retaining nearly all potential planet candidates. Our model correctly identified 10\% to 37\% of the \textcolor{black}{true contaminants}, with a median identification rate of 18\% across all 18 test sectors. Notably, in 16 of the 18 sectors, no planet candidates were classified as contamination. In sector 53, two planet candidates were erroneously flagged as contaminants, while in sector 58, one planet candidate was misclassified. Our contaminant sample predominantly comprised eclipsing binaries, with a median percentage of 82\%.

\textcolor{black}{\subsection{Challenges and Limitations}}

Despite our efforts to avoid incorrectly flagging planet candidates, three were misclassified by our model. All three are \textcolor{black}{TESS Objects of Interest} (TOIs), two of them (TIC 243335710, TIC 372207084) have very short periods of 14.7 and 1.2 days respectively, and the other (TIC 219462190) has a low signal-to-noise ratio (SNR) of 7. \textcolor{black}{This suggests that short-period and low SNR planet candidates may be at higher risk of being incorrectly labeled by our model. However, the periods and SNRs are only available for candidates designated as TOIs. Thus, we were unable to perform a meaningful analysis across the entire dataset to formally test whether the planet's SNR or period relates to its likelihood of being incorrectly flagged by our model.}

Moreover, there is some uncertainty in our labels, \textcolor{black}{which may be introducing some error to our model. The subjective nature of light curve labels and the occasional incomplete marking of transits by citizen scientists contribute to potential gaps in the training/validation data, potentially leading to biased or incomplete models.} To mitigate this, project scientists could manually add transit marks to light curves before training and validating the model. This approach could improve model performance and its ability to generalize to unseen data. However, it is essential to consider the resource-intensive nature of manual vetting, balancing the benefits of improved model performance against the costs and practical constraints associated with the process. Despite the ambiguity in what we interpret as `ground truth' in our training/validation, \texttt{NotPlaNET} can be a helpful tool to reduce the number of light curves requiring manual classification.

\textcolor{black}{\subsection{Comparative Analysis}}

\textcolor{black}{The primary goal of \texttt{NotPlaNET} is to effectively differentiate between real transit events and false positives in photometric time series data from the \textsl{TESS} mission. While injection and recovery tests using synthetic data are common practice in testing models to identify planetary signals, the complexity and variety of false positive scenarios encountered in real \textsl{TESS} light curves is challenging to capture using simulations. These \textcolor{black}{contaminants} can include artifacts, asteroid crossings, background events, and satellite systematics, which are difficult to simulate accurately. }

\textcolor{black}{Given the limitations of simulated data in replicating these diverse scenarios, training or testing the model using a potentially limited set of simulations runs the risk of introducing biases. Instead, we have focused on training and validating the model with real \textsl{TESS} light curves identified by \textsl{Planet Hunters TESS} citizen scientists and confirmed by project scientists.}

\textcolor{black}{As shown in Figure \ref{fig:roc}, the fraction of recovered contaminants is notably lower for Sector 42 compared to the other test sectors. This may be due to Sector 42 containing the first \textit{TESS} observation consisting primarily of the ecliptic plane, and as such contains an increased number of contaminants and a higher level of scattered light.}

\textcolor{black}{\subsection{Future Directions}}

\textcolor{black}{A follow-up study could assess whether model performance improves if employed earlier in the vetting process.} \texttt{NotPlaNET} was trained on and applied to light curves at the end of the \textsl{Planet Hunters} pipeline, where differentiating between a planet candidate and a false positive becomes increasingly difficult. Thus, we suspect our model could also be applied to light curves earlier in the classification process with potentially better results. Employing \texttt{NotPlaNET} in different stages of the vetting process while probing a variety of classification thresholds is a possible direction for future research.

Additionally, our model could be adapted for other science goals by revisiting the scoring method and threshold selection. An alternative method could be to create a scoring system that rewards or penalizes finding contaminants and removing planet candidates differently, depending on the science goals for which the model is being applied. The score $S$ can be written in the form:
\begin{equation}
    S = a N_c - b  N_p ~, 
\end{equation}
where $N_c$ is the number of removed contaminants, $N_p$ is the number of removed planets, and $a$ and $b$ are weighing constants. 

\texttt{NotPlaNET} will be incorporated into the late stages of the \textsl{Planet Hunters TESS} pipeline. After the citizen scientist vetting process described in section \ref{sec:data}, a neural network designed by \citet{malik2022} (\texttt{PlaNET}) will be employed to calculate a score for the vetted light curves and rank them accordingly. Our network will be applied after \texttt{PlaNET} to remove light curves that are confidently contaminants before the project scientist vetting stage. 

\

\section{Conclusion}\label{sec:conc}

We designed \texttt{NotPlaNET}, a CNN trained to identify contaminants from \textsl{TESS} light curves to reduce the number of light curves requiring human vetting. Our model was able to identify a maximum of 37\% and a minimum of 10\% of the contaminants across the 18 test sectors, with a median percentage of 18\%. None of the planet candidates for 16 of the 18 test sectors were incorrectly flagged. For sector 53, 2 planets (0.6\%) were flagged as contaminants. In sector 58, one planet candidate (0.3\%) was incorrectly classified as a contaminant. Moreover, we found our contaminant sample to have a median eclipsing binary percentage of 82\%.

The described methodology has shown potential to help in the identification of exoplanet candidates in large surveys. Our model can be applied to almost any light curve dataset with a variety of science goals. Modifying \texttt{NotPlaNET} and the classification thresholds to accommodate light curves of other dispositions or to find other kinds of astronomical objects like eclipsing binary stars is a potentially interesting follow-up project. Nevertheless, when employed to remove contaminants from \textsl{TESS} data as described in this paper, our model can significantly help reduce the number of light curves that require manual vetting by project scientists and potentially increase our chances of finding and studying interesting long-period planets. 

Although human vetting is, as of now, indispensable in the classification and identification of transiting exoplanets, machine learning continues to show great potential to help speed up the process. Integrating machine learning methods in the vetting pipeline can allow project and citizen scientists to more efficiently parse the data from today's large-scale astronomical surveys. 

\

It is a pleasure to thank
  Shirley Ho (Flatiron),
  Artemis (cat),
  and the members of the Blanton--Hogg group meeting at NYU
for valuable advice and input. Special thanks to Cole for his healthy skepticism, which motivated VTP to prove him wrong and finish this paper.

This paper includes data collected by the \textsl{TESS} mission. Funding for the \textsl{TESS} mission is provided by the NASA Science Mission Directorate. This publication also uses data generated via the Zooniverse.org platform, the development of which is funded by generous support, including a Global Impact Award from Google, and a grant from the Alfred P. Sloan Foundation.

The Flatiron Institute is a division of the Simons Foundation.

\facilities{TESS, MAST}

\software{NumPy \citep{np}, Matplotlib \citep{plt}, PyTorch \citep{pytorch}, scikit-learn \citep{scikit}}

\bibliography{references}{}
\bibliographystyle{aasjournal}

\end{document}